\documentstyle[preprint,aps]{revtex}
\begin{document}
\draft
\title{DOMAIN STRUCTURES OF SMECTIC FILMS FORMED BY BENT-SHAPED MOLECULES }
\author{
E.I.Kats \dag \, \ddag , \, J.Lajzerowicz \dag }

\address{\dag
Laboratoire de Spectrometrie Physique, Univ. Joseph-Fourier Grenoble 1 ,\\
BP87-Saint-Martine D'Heres, Cedex, France, \\
\ddag Landau Institute for Theoretical Physics, RAS \\
117940, GSP-1, Kosygin st., 2, Moscow, Russia.
}
\date{\today}
\maketitle

\begin{abstract}

We formulate a simple Landau type model describing macroscopic
behaviour recently discovered new smectic phases composed of
achiral bent-shaped molecules. Films of such smectics exhibit
three types of ordering related to dipole polarization,
molecular tilt, and chirality. However due to specific third order coupling
of the order parameters these three types of symmetry-breaking
are not independent ones, and this fact leads to specific domain
structures really observed in experiments.
\end{abstract}

\pacs{ PACS number(s): 61.30.Eb, 64.70.Md, 68.10.cr}

\section{INTRODUCTION}

A variety of molecules form liquid crystalline phases (see e.g.
the monograph \cite{CH77}). Many mesogen molecules have symmetries
consistent with the formation of ferroelectric phases and nonzero
dipole moments. Ferroelectric ordering is, however, extremely rare
in positionally disordered liquids or liquid crystals, and since the
discovery of ferroelectric liquid crystals \cite{ML75} it has been
assumed usually that ferroelectricity
is possible only in the chiral smectic -$C^*$ phase (Sm$C^*$),
(formed by chiral molecules) that has the
polar symmetry group $C_2$. In this case polarization can be written
as ${\bf P} = P[{\bf n} \times {\bf z}]$, where ${\bf n}$ is director
\cite{GE74} and ${\bf z}$ is the smectic layer normal.
The necessary conditions for the existence of nonzero polarization are
a finite tilt angle ($\theta \neq 0$) and a molecular dipole
perpendicular to the long axis of molecules.
In racemic mixtures, which contain both enantiomers (that is, molecules
that are mirror images of each other) in equal amounts, the electric
polarization vanishes. Obviously, the electric polarization is directly
connected with molecular chirality in the Sm$C^*$ ferroelectric liquid
crystals.

However there is no fundamental reason that non-chiral liquid crystals
should not be ferroelectrics, since there is
no unambiguous correspondence between chirality of
molecules and the existence of macroscopic ferroelectric properties
or structures they formed.
The attempts of observation of ferroelectricity in non-chiral
liquid crystals are, as a rule, centred around synthesis and investigations
of non-conventional liquid crystalline structures \cite{BL98}.
Recently ferroelectric phases composed of achiral molecules were reported
and investigated \cite{LN97}, \cite{AD99}, \cite{HJ99}.
In these papers it was demonstrated that tilted smectic phases of achiral
molecules show ferroelectric switching, and specific chiral
domain structures. In the paper \cite{BC98} the bulk macroscopic
properties of the lowest possible symmetry smectic phase (triclinic)
were investigated and it was shown that such a system (though formed from
achiral molecules) may possess ferroelectric and piezoelectric
properties as well as macroscopic chirality. Due to polarity within
smectic layers such a smectic may have only integer strength of
point like defects in layers.

Note that in the mentioned above papers (\cite{LN97}, \cite{AD99},
\cite{HJ99}) were investigated only thin freely suspended films
and care must be taken in drawing conclusions about the bulk
properties of liquid crystals from the behaviour of films,
as the surface layers of the film may be in a phase with higher (or lower)
order than the bulk system. The surface phases may not even exist
as bulk phases. In particularly in the papers \cite{LN97}, \cite{HJ99}
were observed not point-like defects, predicted theoretically
in \cite{BC98} for the bulk phase but domain walls, i.e. two-dimensional
defects in smectic layers.

The organization of our paper is the following.
In the next section (II)
we formulate our model and introduce (in the frame work of the Landau
theory) the basic thermodynamics necessary for our discussions.
In the section III we discuss different types of domain structures which
may appear in smectics under consideration, and inspected the role
of external influences (electric or magnetic fields and concentration
of chiral impurities).
The last section (IV)
is devoted to a discussion and summary of our main results.

\section{THEORETICAL MODEL}
According to experimental data presented in the papers \cite{LN97}
and \cite{HJ99}, new smectic structures (labelled in these papers as smectics
$B_2$), are formed by polar but achiral molecules ("banana"-shaped)
having the symmetry group $C_{2v}$, and macroscopic behaviour of these structures is
characterized by three spontaneous symmetry-breaking
leading to the appearance of following properties:
molecular tilt, ferroelectric polarization, and chirality.
The maximal point symmetry group allowing these three types of symmetry
breaking is $C_2$, where the second order axis should be parallel to
smectic planes.

The tilt order parameter in any tilted smectic phases can be characterized
by the two-component order parameter $\psi = \theta exp(i\phi )$,
where $\theta $ is the polar angle (tilt) and $\phi $ is the azimuthal
angle of the nematic director ${\bf n}$.
Instead of $\psi $ one can use so-called ${\bf c}$-director, which is
the projection of the director ${\bf n}$ onto the layer plane.
The magnitude of the tilt order parameter $|{\bf c}| = sin \theta $.
The ferroelectric  polarization ${\bf P}$ is also a vectorial quantity,
and it is only possible along the symmetry axis $C_2$.
From the general point of view the chirality of the system is a third order
antisymmetric tensor which can be reduced for the system under study
to the pseudo-scalar $\chi $.
However in fact we have the only symmetry-breaking, namely
$C_{2v} \to C_2$ and therefore all three order parameters
should be interrelated, and the problem we face now is to find this relation.
In fact since the bend of ${\bf c}$ removes the ${\bf c} - {\bf z}$
mirror symmetry plane, it produces a local chiral symmetry
breaking.
This breaking of chiral symmetry can occur on two distinct length
scales (microscopic or macroscopic). The distinction
between microscopic and macroscopic chiral symmetry breaking
is similar to the distinction between spontaneous and induced
order parameters. From the macroscopic symmetry point of view
to describe chiral, tilted, ferroelectric smectic films we
have to introduce three order parameters $(\chi , \, {\bf c} , \,
{\bf P})$ with a third order coupling $(\chi {\bf c} {\bf P})$
between them.
However in this paper (unlike e.g. \cite{AD99}) we are interested in mainly
microscopic causes of macroscopic symmetry breaking.

From the microscopic viewpoint the existence of a tilt in smectic
phases comes from the requirement of the molecular packing (i.e.
steric forces). These requirements fix for the polar molecules
in our case (thin free standing films) only the module
of the ${\bf c}$-director, and therefore there are two allowed
values of molecular tilt $\pm \theta $.
Thus any molecule in a smectic layer $i$ \footnote{
For the simplicity and according to the layer structure
of smectics, we suppose that the order parameters are uniform
within smectic layers.} can be framed by two state
system labelled by indexes $\pm $ according to the sign
of its tilt. The same manner the dipole moment ${\bf P}$ can be oriented
either parallel or anti-parallel to the second order symmetry axis
and it gives two more states attached to each molecular site.
Therefore each molecular site is a four state system:
$(+ , +) , (+ , -) , (- , +) , (- , -)$, where the first sign
corresponds to the tilt, and the second one to the dipole moment.
If among the $N^i$ molecules
in a certain smectic layer $i$
the number of molecules in each state is $N^i(+ , +) , N^i(+ , -) ,
N^i(- , +) , N^i(- , -)$
then evidently
\begin{eqnarray}
\label{j1}
N^i = N^i(+ , +) + N^i(+ , -) + N^i(- , +) + N^i(- , -)
\end{eqnarray}

Analogously it is easy to see, that the tilt angle for the layer $i$ can be
represented as:

\begin{eqnarray}
\label{j2}
N^i \theta ^i = N^i(+ , +) + N^i(+ , -) - N^i(- , +) - N^i(- , -) \, ,
\end{eqnarray}

and the polarization is given by:
\begin{eqnarray}
\label{j3}
N^i P^i = N^i(+ , +) + N^i(- , +) - N^i(+ , -) - N^i(- , -)
\end{eqnarray}

It is important to note that for each molecular site the product of
$P \theta $ represents the chirality of the given molecule, independently of
site and of layer $i$. We follow here the idea and method developed
recently for solid racemic solutions \cite{LL91}.
However though for each individual molecular site $\chi \equiv P \theta $,
this relation generally is not valid for the local mean values for a layer
$i$, i.e. $\theta ^i P^i \neq \chi ^i$, since analogously to (\ref{j2},
\ref{j3}) one can write:

\begin{eqnarray}
\label{j4}
N^i \chi ^i = N^i(+ , +) + N^i(- , -) - N^i(- , +) - N^i(+ , -) \, .
\end{eqnarray}
In the spirit
of the Bragg Williams mean field approximation we can compute the
entropy of the system
\begin{eqnarray}
\label{j5}
S = ln \left [\frac{N!}{N(+ , +)!N(+ , -)!N(- , +)!N(- , -)!}  \right ]\, ,
\end{eqnarray}
where $N = \sum _{i} N^i$ the total number of molecules.

Solving the equations (\ref{j1} - \ref{j4}), introducing the found
expressions for $N^i(\pm , \pm )$ in terms of the order parameters
$\theta , P , \chi $, and expanding of (\ref{j5}) for small values
of the order parameters we get
$$
S = - N[ \frac{1}{2}(P^2 + \theta ^2 + \chi ^2) + \frac{1}{2}(P^2 \chi ^2
+ P^2 \theta ^2 + \chi ^2 \theta ^2) + \frac{1}{12}(P^4
+ \chi ^4 + \theta ^4) - \theta \chi P]
$$
It is important to notice (and this is one of the main points of the our
investigation) the presence of the specific third order term $\theta P \chi $.
The free energy of the system $F = U - T S$ (where $U$ is the internal
energy associated to intermolecular interactions) should have the same
structure as the entropy $S$ but with renormalized coefficients, namely
\begin{eqnarray}
\label{j6}
F = \frac{a_1}{2}P^2 + \frac{a_2}{2}\theta ^2 + \frac{a_3}{2}\chi ^2 +
\frac{b_1}{2}P^2\chi ^2 + \frac{b_2}{2}\theta ^2P^2 + \frac{b_3}{2}\chi ^2
\theta ^2 + \frac{c_1}{2}P^4 + \frac{c_2}{2}\theta ^4 + \frac{c_3}{2}\chi ^4
+ \gamma \chi \theta P
\end{eqnarray}
The fact that the third order term necessarily figures in the free energy
does not change at the renormalization and it is related to the symmetry,
since the product of the three representations to which
$\theta  , \chi  , $ and $P $ belong includes the identical representation.

The coefficients $a_i , b_i , c_i$, and $\gamma $ can be considered as
phenomenological parameters and $a_i$ should become small near the
corresponding symmetry-breaking transitions.
To say more requires further knowledge of all these coefficients.
Unfortunately using only the data known from the literature we are not
able to extract values of all needed parameters. Therefore we will not compare
quantitatively our theory with available experimental data, since
with too many unknown parameters the theory tends to become an exercise
in curve fitting, which looses predictive credibility. Instead of this we will
discuss in the next section qualitative features of the model.

\section{Qualitative analysis of the model}

Let us consider some very general consequences of the model.
First note that the third order coupling found above corresponds to
account of three particle interactions. If we suppose to escape a
conflict between experiment and theory that all three order parameters
are uniform within smectic planes, this third order coupling
means that the modulations of
the order parameters along the normal to smectic layers should be matched
$\chi (q_1) \theta (q_2)  P(q_3)$ to provide $q_1 +q_2 +q_3=0$.
Due to smectic periodicity along this direction $|q_i| \equiv q_0$, where
$q_0 = 2\pi /d$ is the wave vector of smectic density modulation ($d$ is the
interlayer distance). Thus to satisfy the matching there are only two
possibilities: (1) one of the three wave vectors is zero and two others
are anti-parallel; (2) all three wave vectors are zero.

Second let us assume that one from the three coefficients $a_i$
is much smaller than two others. Therefore in the temperature region
where this condition is fulfilled, we have only one soft order
parameter, and one may neglect two others (hard)
degrees of freedom.
In this case the theory is reduced to the well known Landau
theory for a scalar order parameter \cite{BI86}. However due to
its importance for the present context (and for convenience) we repeat
mainly known results to apply them
to our concrete case (free standing films).
It is just the case where it is easy and more useful to derive
these results for the concrete system under consideration
than to try to find the suitable references, and to modify
all expressions to apply them to the case.

There are two effects, related to the existence of the surface in free
standing films. The first is a pure geometrical one (finite size
effects). The surfaces break
the translational
and rotational invariance (because the surface
is a specific plane which breaks the translational
invariance, and the normal to the surface is a specific direction which
breaks the rotational invariance). Besides, certainly, there are
physical modifications of the system due to the existence of the surface
(surface effects).
The surface can suppress the bulk ordering (this case is traditionally
called the ordinary phase transition), the surface can enhance the
bulk ordering (it is called the extraordinary phase transition),
or as a third possibility the surface can experience its
intrinsic critical behaviour. There is also a so - called special
phase transition which is intermediate between ordinary and
extraordinary transitions.

The both effects
related to the existence of the surface
can be taken into consideration in the frame work
of the Landau expansion. In our particular case (film geometry and
$a_1 \ll a_2 , a_3$) it has the form:

\begin{eqnarray}
\label{m1}
F = \int _{0}^{L} dz \biggl (\frac{1}{2} a_1 \theta ^2
+\frac{1}{4} c_1\theta ^4  + \frac{1}{2} d_1(\nabla \theta )^2 \biggr ) +F_s
\end{eqnarray}
where we added to (\ref{m1}) the gradient term (with the coefficient $d_1$) to
describe the tilt profile over the film thickness $L$, and $F_s$
is the surface energy which should have the same form as the
bulk energy (\ref{m1}):
\begin{eqnarray}
\label{m2}
F_s = \frac{a'}{2}\left(\theta ^2(0)+\theta ^2(L)\right)+\frac{c'}{4}
\left(\theta ^4(0)+ \theta ^4(L)\right)
\end{eqnarray}
Usually it is supposed that $a' \equiv d \lambda ^{-1}$, where
$\lambda $ is called by extrapolation length and experimental data indicate
that (at least as it concerns to the tilt) we have $\lambda < 0$
and it is called traditionally by extraordinary phase transition.

In this case the surface enhances the ordering and therefore
on the surface one can expect the onset of ordering before (i.e.
at higher temperatures) it occurs in the bulk.
So one can expect in this case the surface transition for temperatures
$T_s > T_c$ (by the definition the bulk transition temperature
is determined from $a_1(T_c) = 0$). But of course at $T_c$
due to the onset of the bulk order the surface will experience
some critical behaviour as well. In the regime of $T_c < T < T_s$
the bulk correlation length $\xi _b$ is finite and the order parameter
decays from its maximum value at the surface.
One can easily find the transition temperature for the surface
layer:
$$
\frac{T_s}{T_c} - 1 = \frac{d_1}{T_c} \lambda ^{-2}
$$
To find the profile for the order parameter we have to solve
the Euler -  Lagrange equation which follows from (\ref{m1})
supplemented by the boundary condition, which can be found from (\ref{m2}).
Performing this rather routine procedure one can find that there are
two types of configurations providing the minimum of the bulk functional
(\ref{m1}) and simultaneously minimizing the surface energy (\ref{m2}).
The first let say natural solution is symmetrical one (we will term this
solution by synclinic structure):

${\bf S-configuration:} \, \theta (z= 0) = \theta (z=L)$,

We imply that $a'=\alpha'(T-T_s)$ where $T_s$ is the surface transition temperature
(it can be extracted from experimental
data for very thin films, e.g. for two-layer films).
Determining
the surface transition temperature we can omit third-order term in the
equation for the bulk, and the transition
in the film with $N$-layers occurs at $T_N$ which can be found from
the following equation
$$
T_N = T_s - \frac{d_1}{\alpha'\xi_b(T_N)}
\tanh\left(\frac{L}{2\xi_b(T_N)}\right)
$$

The second solution (we will term it by anticlinic) is antisymmetrical one:

${\bf A-configuration:} \,  \theta (z=0) =-\theta (z=L)
$
and for this case
$$
T_N = T_s - \frac{d_1}{\alpha'\xi_b(T_N)}
\coth\left(\frac{L}{2\xi_b(T_N)}\right)
$$

The solution of both transcendental equations can be found numerically
very easily and (as it should be) for small film thicknesses
$L \ll \xi _b(T_N)$ synclinic configuration has always the higher
transition temperature while for thick films with $L \gg \xi _b(T_N)$
the anticlinic solution can have the higher transition temperature.
But certainly the anticlinic state can be only metastable one due to
the gradient energy (or by the other words due to the energy
penalty which one must pay for the domain wall appearing inevitably
for the anticlinic structure).
However the given above statement is valid only for the case
$a_1 \ll a_2 , a_3$, when we have deal with one scalar order
parameter condensation. It is not the case when we have two or three
soft degrees of freedom (condensed order parameters) due to third
order coupling between them.

As we have seen already the minimization of the third order coupling energy
(in the conditions when all three order parameters are condensed) leads
to the following possible structures of smectics under consideration:

(i)
$$\theta (q=0) ;\, P(q=q_0) ; \, \chi (q=-q_0) $$
i.e. synclinic, antiferroelectric and racemic;

(ii)
$$ \theta (q=q_0) P(q=0) \chi (q =-q_0) $$
i.e. anticlinic, ferroelectric and racemic;

(iii)
$$
\theta (q=q_0) P(q=-q_0) \chi (q=0)
$$
i.e. anticlinic, antiferroelectric and homochiral;

(iv)
$$
\theta (q=0) P(q=0) \chi (q=0)
$$
i.e. synclinic, ferroelectric and homochiral.

It is worth to note that all four types of predicted structures
are really observed in experiments \cite{LN97} , \cite{HJ99}.
Even more, it is clear that the application of the external
electric field should stimulate the ferroelectric ordering
of dipoles and therefore only (ii) and (iv) structures
will be stable in a strong enough field. And it is also
exactly what was observed in \cite{LN97}.
The same manner the external field conjugated to the chirality
should induce (iii) and (iv) structures. As a physical realization of
this field one can have in mind the concentration of homochiral impurities.
And the field conjugated to the tilt angle must induce
(i) and (iv) structures only. Physically such a field can be provided
by the anchoring.

In the case when we have the condensation of two order parameters
(there are three types of such pairs) one can observed a very rich
behaviour with many types of domain walls. For each type of the wall
at the variation of the parameters $a_1 , a_2 , a_3$
the wall transformations can be observed, which can be understood as an Ising - Bloch
phase transitions with the domain wall symmetry breaking.
Unfortunately, we can find no guidance from experimental or theoretical
sources for choosing all phenomenological coefficients that appear
in these expressions. Thus the primary function of this section
must be to give a qualitative interpretation of our results
and to demonstrate the possibility of ferroelectric ordering in basically
non - chiral systems, as opposed to proving exactly its existence.

\section{Conclusion}

We formulated a simple Landau type model describing macroscopic
behaviour recently discovered new smectic phases composed of
achiral bent-shaped molecules. Films of such smectics exhibit
three types of ordering related to dipole polarization,
molecular tilt, and chirality. However due to specific third order coupling
of the order parameters these three types of symmetry-breaking
are not independent ones, and this fact leads to specific
structures (i) - (iv) really observed in experiments.
This inhomogeneous ordering physically means that over a
large region of thicknesses of free standing films they can be
considered as some effective interfaces.
It is typical for liquid crystals
\cite{FP84}
that the width of the interface of experimental mesogenes is 40 - 100
times the length of molecules. We observed the example of how the
presence of an interface may induce a type of ordering in the
inhomogeneous region (for free standing films it may be the whole thickness
of the system) that does not occur in the bulk phases.
The analogous phenomena are known also for Langmuir monolayers where
chiral symmetry can be spontaneously broken \cite{SW93}, and it leads
to a chiral phase composed of non - chiral molecules.
In fact for a thick free standing film the top and bottom layers
are each equivalent to Langmuir monolayers.

The tilt arrangement
of the $A$ configuration is anticlinic, i.e. the top and the bottom of the
film are tilted in opposite directions.\footnote{
Note also recent ellipsometric studies \cite{SB98}
where new phases of liquid crystals have been investigated
and among these are the antiferroelectric Sm$C_A$ structures where the
tilt direction alternates when going from layer
to layer.}

Physical mechanisms providing the
polarization properties of non-chiral and chiral free standing
films are very different. For the non-chiral systems the polar order
is induced in fact by the steric packing of anisotropic (but non-chiral)
molecules, whereas in the ordinary (chiral) ferroelectric
liquid crystalline phases the polar order is a consequence of
the molecular chirality.

\acknowledgements
This work was supported in part by RFFR and INTAS grants
and  by
the Russian State Program "Statistical Physics".
E.K. thanks prof. M. Vallade
for supporting his stay at the Lab. Spectro., Joseph - Fourier
University Grenoble - 1 and
for fruitful discussions.

\end{document}